# Absolute bacterial cell enumeration using flow cytometry

[Bacterial cell enumeration using flow cytometry]

F. Ou[1], C. McGoverin[1], S. Swift[2], F. Vanholsbeeck[1]

[1] The Dodd-Walls Centre for Photonic and Quantum Technologies, Department of Physics, The University of Auckland, Auckland, New Zealand.

[2] School of Medical Sciences, The University of Auckland, Auckland, New Zealand.

**Correspondence**

Frédérique Vanholsbeeck, The Dodd-Walls Centre for Photonic and Quantum Technologies, Department of Physics, The University of Auckland, Auckland, New Zealand.

E-mail: f.vanholsbeeck@auckland.ac.nz

## Abstract

**Aim**: To evaluate a flow cytometry protocol that uses reference beads for the enumeration of live and dead bacteria present in a mixture.

**Methods and Results:** Mixtures of live and dead *Escherichia coli* with live:dead concentration ratios varying from 0 to 100% were prepared. These samples were stained using SYTO 9 and propidium iodide and 6 µm reference beads were added. Bacteria present in live samples were enumerated by agar plate counting. Bacteria present in dead samples were enumerated by agar plate counting before treatment with isopropanol. There is a linear relationship between the presented flow cytometry method and agar plate counts for live ($R^2 = 0.99$) and dead *E. coli* ($R^2 = 0.93$) concentrations of *ca.* $10^4$ to $10^8$ bacteria ml$^{-1}$ within mixtures of live and dead bacteria.

**Conclusions:** Reliable enumeration of live *E. coli* within a mixture of both live and dead was possible for concentration ratios of above 2.5% live and for the enumeration of dead *E. coli* the lower limit was *ca.* 20% dead.

**Significance and Impact of the Study:** The ability to obtain absolute cell concentrations is only available for selected flow cytometers, this study describes a method for accurate enumeration that is applicable to basic flow cytometers without specialised counting features. By demonstrating the application of the method to count *E. coli,* we raised points of consideration for using this FCM counting method and aim to lay the foundation for future work that uses similar methods for different bacterial strains.

**Keywords**

Fluorescence, Bacterial cell enumeration, Flow cytometry, Counting beads, Detection, Bacteria

# Introduction

The accurate determination of both live and dead bacteria is important in many applications, ranging from monitoring bactericidal efficacy to optimisation of industrial fermentation processes. The agar plate count or viable count method is the established standard reference method for the enumeration of live bacteria, and is routinely applied in clinical, industrial and research settings. Agar plate counting allows determination of bacterial concentrations via counts of colony forming units (CFU), with the assumption that each CFU grew from one bacterium of the sample. However, often bacterial aggregates are not broken up and the cells are not evenly dispersed on the plate. As it is likely that each CFU arose from more than one bacterium, the resulting CFU count will underestimate the actual number of viable bacteria present (Daley, R. J., 1979; Jansson and Prosser, 1997; Auty et al., 2001). Agar plate counting only enumerates the cells that are culturable under the conditions of the investigation and cannot count dead cells or the viable but non-culturable (VBNC) cells, i.e. cells that retain cellular and metabolic activity but are stressed (Bensch et al., 2014). Moreover, agar plate counting is a time-consuming, multi-day process, thus it does not provide timely information that is required in applications such as industrial manufacturing, research and medical diagnoses.

The detection of VBNC bacteria can be achieved using fluorescence stains that target and label specific cellular properties such as membrane potential, esterase activity and dehydrogenase activity (Vives-Rego et al., 2000). The cellular property of interest in this study is membrane integrity, which is highly ranked in the assessment of physiological state (Vives-Rego et al., 2000). Fluorescence detection is achieved mainly by microscopy, a technique that is labour- and time- intensive despite the continued developments in apparatus automation, image analysis and dye-specificity (Hammes et al., 2008; Khan et al., 2010). Due to the statistical counting error ($n^{0.5}$), a large number of events must be investigated to minimise uncertainties of the counts (Nebe-von-Caron et al., 2000). For example, in order to obtain a statistical coefficient of variation below 3%, 1000 events need to be investigated (Nebe-von-Caron et al., 2000), which is laborious using either microscopy or plate count methods. Image cytometry is a related technique in which the system is capable of automatically obtaining numerous static images and counts of different cell populations. However its application to microbiology is limited due to an inability to resolve the small size of bacterial cells (ThermoFisher Scientific, 2013; Nexcelom Biosciences, 2017).

Over the last few decades, flow cytometry (FCM) has become an increasingly important tool for microbiologists to study cells at both the individual and the population levels. FCM is a reliable technique based on the optical detection of scattered light and fluorescence that allows the identification of cells with particular characteristics of interest (e.g. intact cytoplasmic membrane). FCM is fast becoming the preferred choice for obtaining rapid and multi-parametric information at the single-cell level (Yang et al., 2010; Van Nevel et al., 2017b). It was first applied to measure the nucleic acid content and light scattering of unstained mammalian cells in the 1960's (Kamentsky et al., 1965). Since then, the development of optics technology and availability of specific bacterial fluorescent dyes have made it possible to sensitively detect bacterial cells using FCM (Davey and Kell, 1996; Tracy et al., 2010; Yang et al., 2010).

When a sample is stained with dyes that differentially bind to live and dead bacteria, FCM measurements may be used to identify live as well as dead bacteria. One of the main advantages of FCM over the traditional plate count and fluorescence microscopy methods is the ability to obtain measurements of single cells in large sample sets with limited effort. One limitation common to plate counts, microscopy and FCM methods is cell aggregation which will result in slight underestimation of cell counts (Gunasekera et al., 2000). In FCM, this is commonly referred to as coincidence and can be minimised by using a sufficiently low flow rate or low concentration of cells (Alsharif and Godfrey, 2002; Gasol and Del Giorgio, 2000).

Despite having the ability to measure the relative proportions of cell populations in a mixture, not all flow cytometers can obtain the absolute concentration of the cells. The reason being many models of flow cytometers have no way of precisely controlling the flow of the sample through its interrogation point (Gasol and Del Giorgio, 2000). Therefore, the number of particles analysed in one cytometric run cannot be immediately correlated to a given sample volume to obtain a measurement of particle density. The easiest, most reliable and inexpensive way of obtaining absolute counts with FCM is to use reference beads (Gasol and Del Giorgio, 2000).

Many bacterial enumeration studies have been published but most have used FCM systems that are equipped with volumetric control to enable direct quantification of bacterial concentration (Buzatu et al., 2014; Bettarel et al., 2016; Carlson-Jones et al., 2016; Frossard et al., 2016; Hildebrandt et al., 2016; Fontana et al., 2017; Nocker et al., 2017; Van Nevel et al., 2017a). These instruments are not available in all laboratories and are expensive. Previous studies have used the bead-based method to correlate FCM counts of live bacteria to plate counts (Alsharif and Godfrey, 2002; He et al., 2017), and hemocytometer counts (Peniuk et al., 2016). However, these studies have not investigated a wide range of concentrations or have not covered a wide range of live:dead concentration ratios. In addition, the enumeration of dead cells have not been validated or presented.

Application of the bead-based FCM counting method to enumerate live and dead bacteria has been proposed by the manufacturer since the release of the Baclight LIVE/DEAD Bacterial Viability and Counting Kit more than one decade ago (ThermoFisher Scientific, 2004). However, a number of precautions of the methodology and analysis of associated experimental errors have not been included. The purpose of this study is to raise points of consideration when using this method of cell counting and aims to lay the foundation for future work that uses this counting method for different bacterial strains. The current study uses the bead-based FCM method for counting *Escherichia coli* over a wide range of total bacterial concentration, and for live:dead *E. coli* concentration ratios ranging from 0% to 100%. We demonstrated reliable enumeration of live *E. coli* when the live:dead concentration ratio ranged from 100% down to *ca.* 2.5% live; and reliable enumeration of dead when the percentage of dead was between 100% and *ca.* 20%. Comparison of FCM and plate counts gave a linear relationship from $10^4$ to $10^8$ bacteria ml$^{-1}$ for live ($R^2 = 0.99$) and for dead *E. coli* ($R^2 = 0.93$). Our bead-based method is applicable to common FCM systems and the rapidity, reliability and low error of the method makes it attractive for enumeration of bacterial populations where little prior knowledge of bacterial cell concentration exists.

## Materials and Methods

### Bacterial growth

As a model, *E. coli* (ATCC 25922; Cryosite, Granville, NSW, Australia) was used for all experiments. *E. coli* was chosen because it is a widely used test organism in microbiology and is also the most thoroughly studied species of bacteria (Cooper, 2000). Live and dead bacteria for FCM were prepared according to the scheme summarised in Fig. 1. Briefly, *E. coli* was incubated overnight in Difco tryptic soy broth (TSB; Fort Richard Laboratories, Auckland, New Zealand) then sub-cultured in fresh TSB (20× dilution) and incubated for approximately 1 h. The sub-culture was grown to reach *ca.* $4\times10^8$ CFU ml$^{-1}$, with an optical density between 0.5 and 0.6 at 600 nm (path length 1cm). All broth cultures were grown at 37°C and aerated with orbital shaking at 200 rpm.

# Preparation of live:dead bacterial mixtures

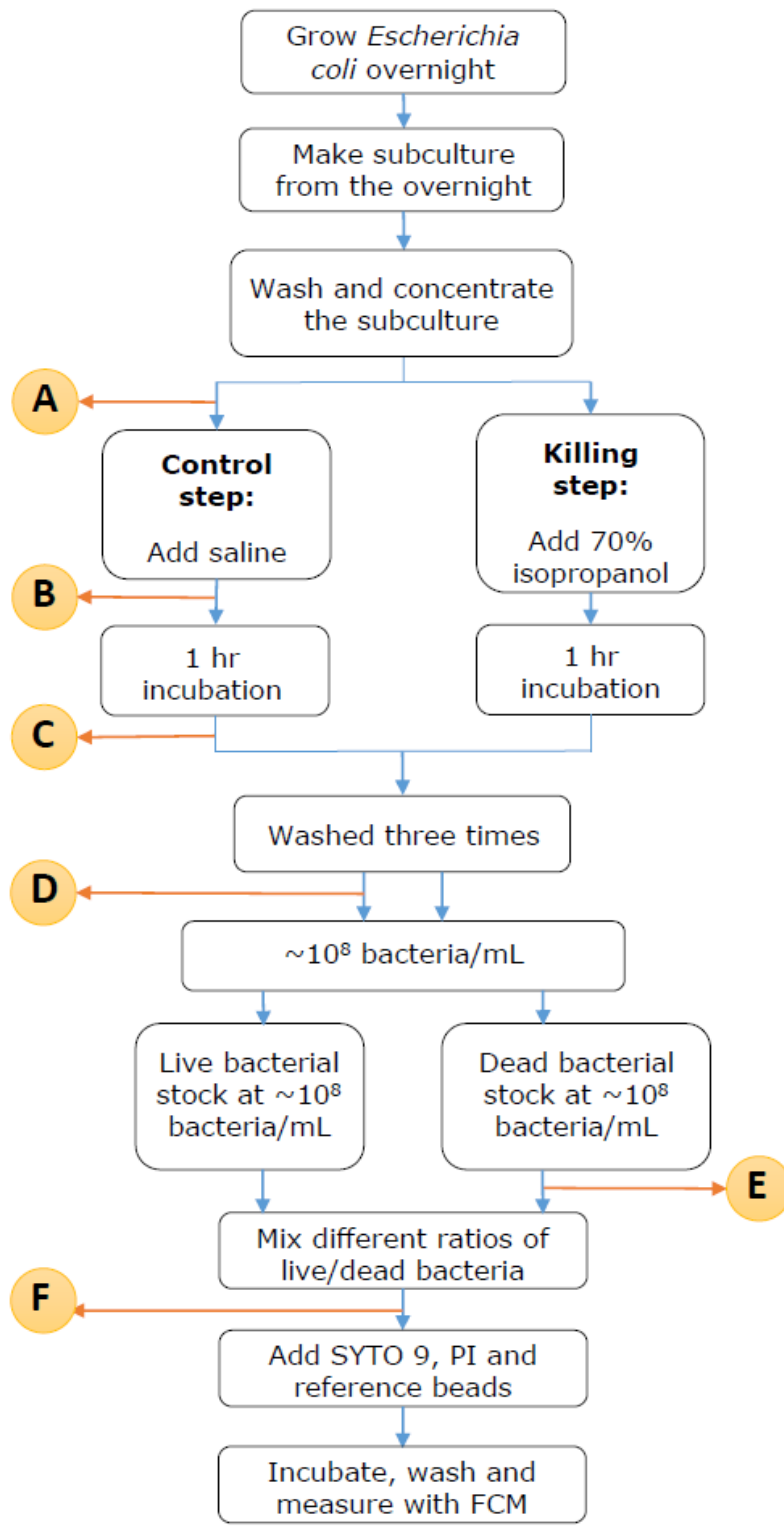

**Figure 1** Work flow diagram for the preparation of samples to be measured by flow cytometry (FCM), the orange circles indicate where plate counts were performed. Plate counts were done for the concentrated sub-

culture (A) and after it was diluted by saline (B), to infer the number of bacteria involved in the killing step. Plate counts were performed at steps C and D to monitor changes in bacterial concentration due to incubation and washing, respectively. The dead bacterial stock was plated to confirm killing (E). To obtain the concentration of live bacteria in the final samples, plate counts were done (F) in parallel with taking the FCM measurements.

Bacterial suspensions were made using a modified protocol based on the instructions from the BacLight LIVE/DEAD Bacterial Viability and Counting Kits manual (ThermoFisher Scientific, 2004). The sample preparation is illustrated in Fig. 1. Exponentially growing cells were harvested by centrifugation ($4302 \times g$, 10 minutes, 21°C) and re-suspended in 3 ml of saline (0.85% w/v) after removal of the supernatant. Subsequently, 1 ml of the washed sub-culture was diluted with either 9 ml of saline (live bacterial solution) or 9 ml of 70% isopropyl alcohol (dead bacterial solution). Each solution was incubated for 1 hour at 28°C and shaken at 200 rpm. Live and dead cells were harvested by centrifugation ($4302 \times g$, 10 minutes, 21°C) and washed three times. During each washing step the pellet was re-suspended in 20 ml saline, then centrifuged at $4302 \times g$ for 10 minutes, at 21°C. After the final wash, the cells were re-suspended in saline to achieve a concentration of *ca*. $1 \times 10^8$ bacteria ml$^{-1}$; equivalent to diluting the sample to an optical density of *ca*. 0.2 at 600 nm. The live and dead bacterial stock solutions were diluted to concentrations ranging from $10^4$ to $10^8$ bacteria ml$^{-1}$. These diluted live and dead bacterial solutions were then combined in various live:dead ratios, giving approximately 0, 2.5, 5, 10, 20, 50, 80 and 100% live bacteria.

## Enumeration via plate count

### Live bacteria

Bacterial samples were diluted to approximately $10^3$ bacteria ml$^{-1}$ and 100 µl was spread on Difco tryptic soy agar plates (Fort Richard Laboratories). Three replicate aliquots were plated from each sample. The saline was plated to check sterility.

## Dead bacteria

The number of live bacteria can be determined via the plate count method, but no information can be directly obtained about the levels of dead bacteria. We estimated the concentration of dead bacteria present in the dead bacterial stock indirectly from plate counts. To achieve this, the concentrated sub-culture and the 10 times diluted sub-culture were plated, steps A and B respectively, as shown in Fig. 1. The average CFU ml$^{-1}$ obtained from steps A and B were used to calculate the 'concentration of bacteria killed' by incubation in 70% isopropanol. The 'expected concentration of dead bacteria' in the final samples were calculated using the formula below.

$$\text{expected conc. of dead bacteria} = \frac{\text{conc. of dead bacteria} \times \text{vol. of dead bacteria added}}{\text{vol. of mixture}} \qquad (1)$$

where the concentration of dead bacteria is the concentration of the dead bacterial stock at approximately $10^8$ bacteria ml$^{-1}$ and is obtained from the plate counts done at steps A and B in Fig. 1. The volume of dead bacteria added is the volume of the dead bacterial stock.

## Fluorescent dye staining and microsphere protocol

Baclight LIVE/DEAD Bacterial Viability and Counting Kits (Invitrogen, Molecular Probes, Carlsbad, CA, USA, L34856) were used in our experiments. The kit consists of a microsphere suspension and two types of fluorescent dyes SYTO 9 and propidium iodide (PI) that label live and dead bacteria, respectively (ThermoFisher Scientific, 2004). Both dyes are stored in DMSO, SYTO 9 at a concentration of 3.34 mmol l$^{-1}$ and PI at 20 mmol l$^{-1}$. The microspheres are suspended in deionised water containing 2mmol l$^{-1}$ sodium azide, at a concentration of $1 \times 10^8$ beads ml$^{-1}$. Each microsphere has a diameter of 6 μm. The microsphere suspension was homogenised by a series of gentle inversions of the bottle followed by sonication in a water bath for 5 – 10 minutes prior to dispensing.

For staining, 1 μl of SYTO 9 and 1 μl of PI were aliquoted into a microcentrifuge tube, followed by 10 μl of the microsphere suspension and then 990 μl of the bacterial suspension. Each sample was incubated in the dark for 15 minutes at room temperature to allow dye-bacteria binding. Immediately before measuring on the flow cytometer, the samples were mixed by gentle inversions and vortexed carefully at 5 rpm (*ca.* 0.00013 × *g*) to minimise foam formation.

## Enumeration using flow cytometry

All samples were evaluated using a LSR II Flow Cytometer (BD biosciences, San Jose, CA, USA). A 488 nm laser with 20 mW power was used for excitation. SYTO 9 fluorescence was collected using a 505 nm longpass filter and bandpass filter with transmission at 530/30 nm. The PI fluorescence was collected using a 685 nm longpass filter and bandpass filter with 695/40 nm transmission. To minimise noise, threshold was set to side scatter (SSC) at 200. The photomultiplier tube voltage was adjusted so that both the bacterial populations and beads were on scale in a SSC vs forward scatter (FSC) plot, and a time histogram was incorporated in the analysis to enable observation of bead count consistency. Analysis of live and dead *E. coli* was done by framing the various populations in the red fluorescence versus green fluorescence cytogram (Fig. S1). In this study, *E. coli* concentrations were kept to a maximum of *ca.* $10^8$ bacteria ml$^{-1}$ to avoid the need for dilutions of the sample. To maintain a constant setting for all measurements while reducing the occurrence of coincident detection of bacteria in the high concentration samples, flow rate was kept to approximately 6 µl min$^{-1}$ and the duration of each measurement was 150 s.

The number of microsphere beads added was used to calculate the absolute concentration of bacteria measured via FCM. The formula below describes the calculation for the concentration of live bacteria, the same logic was used to find the concentration of dead bacteria (ThermoFisher Scientific, 2004; Khan et al., 2010).

$$\text{concentration of live bacteria} = \frac{\text{no. of events in live region}}{\text{no. of events in bead region}} \times \text{conc. of beads} \times \text{dilution factor} \qquad (2)$$

where the *no. of events in live region* represents the cells stained by SYTO 9 but not PI, which excludes dead cells. In equation (2), the *concentration of beads* refers to the bead concentration in the entire sample volume and the *dilution factor* refers to the dilution of the bacterial sample.

To test the variability in the instrument, each sample was divided into three analytical replicates to take triplicate measurements. The variability in the bead addition step was also tested, by pipetting dye and beads to three different tubes before adding the same bacterial suspensions to each tube. Multiple packs of BacLight LIVE/DEAD Bacterial Viability and Counting Kits were used in the FCM-based enumeration experiments, to examine whether bottles of beads from different batches cause variation in bacterial counts.

### Statistical analysis

From triplicate measurements of plate counts, FCM counts and percentage live/dead, the mean and standard errors (SE) were found using the respective formulas in Microsoft Excel. The SE of regression, residual analysis and the $R^2$ for both the live and dead standard curves were found via Python programming using the numpy and maplotlib packages. The SE was used to examine the measurement variation, as 95% of the measurement means are expected to be within 2SE of the expected value.

## Results

### Bacterial concentration variation induced by sample preparation

To assess whether sample preparation steps influenced the number of cells analysed by FCM, we measured (i) the live bacteria that grew during the hour-long incubation; and (ii) the bacteria removed by the washing process (steps C and D respectively, Fig. 1). Triplicate plate counts were done before and after the incubation (Fig. 2a) and washing (Fig. 2b) processes and the CFU ml$^{-1}$ recorded. The difference between the CFU ml$^{-1}$ before and after the respective processes are plotted with their SE. The hour-long incubation is more likely to increase the plate count, whereas the effect of the washing process on the results are variable. The incubation step had a mean absolute percentage deviation of 45%. The washing steps had a mean absolute percentage deviation of 28%.

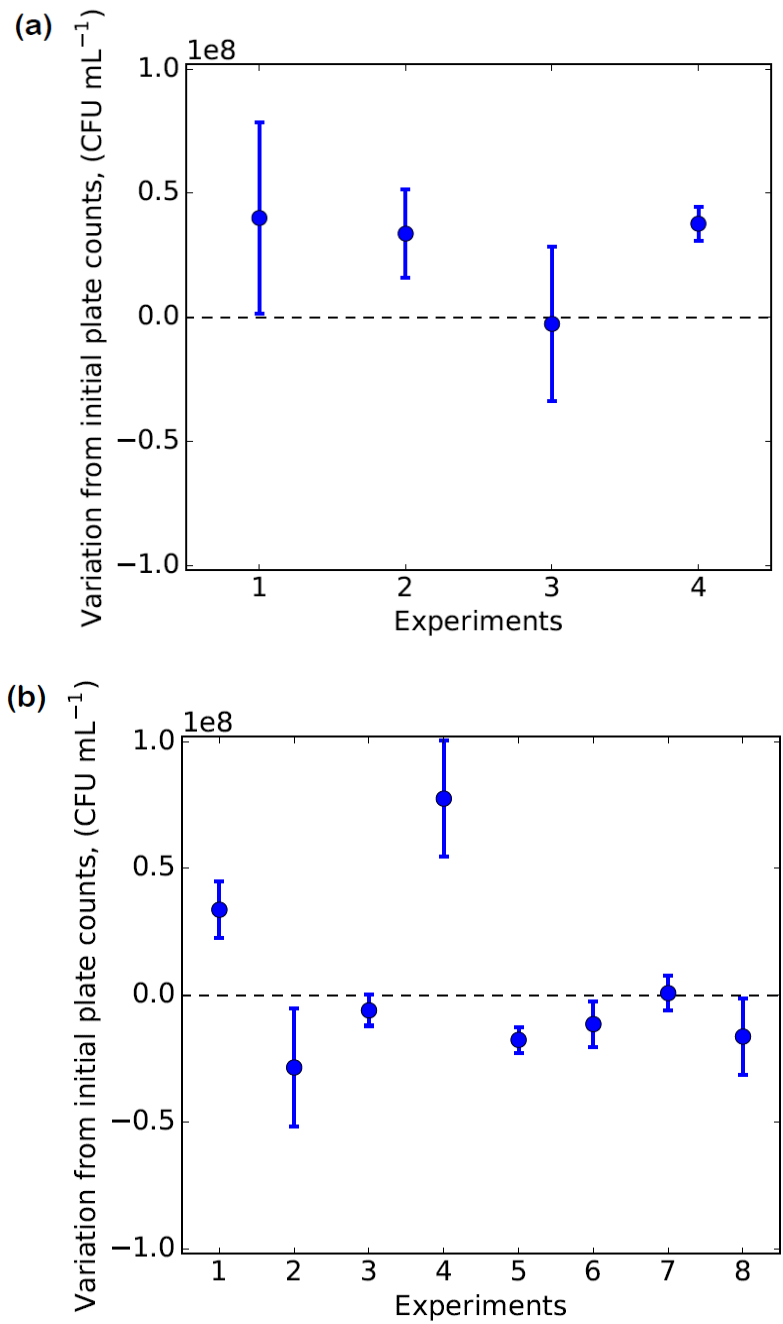

**Figure 2** The effect of sample preparation steps on the concentration of cells in flow cytometry samples. The change in plate counts obtained after the incubation (a) and washing processes (b) of sample preparation, relative to the plate counts obtained before the respective processes. The error bars indicate the standard error in triplicate plate counts.

## Standard curve of live bacteria

To assess whether calibration of the flow cytometer with bottles of beads from different batches affected the bacterial count obtained, different packs of BacLight LIVE/DEAD Bacterial Viability and Counting Kit were used. *E. coli* mixtures with live bacterial concentration ranging from $10^4$ to $10^8$ bacteria ml$^{-1}$ were measured, as shown in Fig. 3. The results show that there were no major variation from using the different bottles of beads.

To validate FCM counts with the established plate count method, the counts obtained from both methods were compared. An overall standard curve for enumerating live bacteria was calculated (y = 0.958x + 0.362), which is significantly different from a 1:1 line. $R^2$ for the overall standard curve of live *E. coli* is 0.99 and the SE of linear regression is 0.121.

The FCM-measured concentrations were consistently higher than those obtained by the plate count method. Triplicate measurements were recorded and the data showed there were no major variations in the instrument and sample preparation. Each data point (N = 70) on Fig. 3 is the mean of triplicate measurements (n ≈ 210). The experiments that obtained these data were repeated on multiple days.

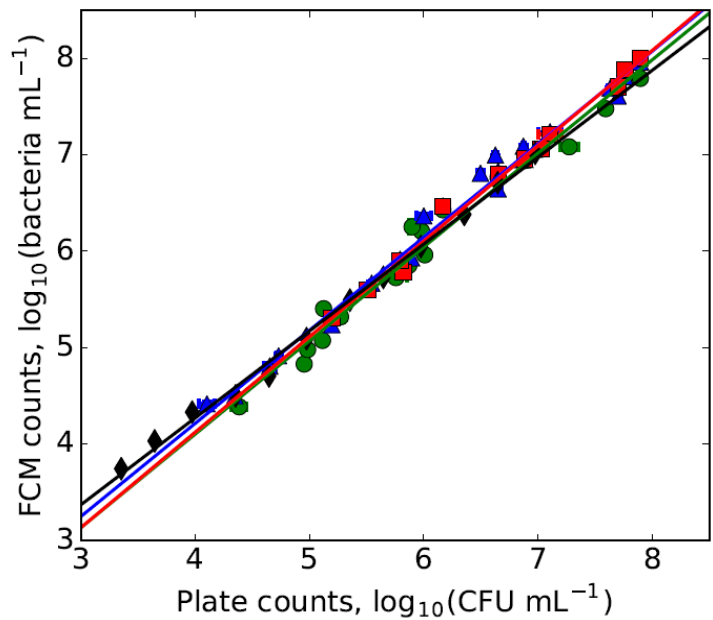

**Figure 3** Log of live *E. coli* concentration obtained by flow cytometry (FCM) was compared to that obtained via plate counts. Data were collected using different bottles of beads, numbered 1 (●), 2 (▲), 3 (■), and 4

(♦).The different standard curves are defined by the different bottle of beads used in the FCM experiment. The error bars in x and y directions indicate the standard error in triplicate samples.

## Validation for the standard curve of live bacteria

To validate the FCM-based enumeration of live bacteria, 31 test set samples were collected over 2 days and using dyes and beads from three different LIVE/DEAD Bacterial Viability and Counting Kits. Fig. 4 shows that with the exception of one outlier, all other test set samples fit reasonably well with the overall standard curve obtained for live *E. coli*, including data obtained using bead bottle 5 which was not used in the establishment of the standard curve.

To check the predictability of the overall standard curve for live bacteria, residual analysis was performed. The analysis of the residuals indicated that bottle 1 residuals were mostly below zero, bottle 3 above zero and bottle 5 both above and below zero, consistent with the skew in Fig. 4. Overall the value of the residuals were fairly constant across the tested concentration range, excluding the outlier. The SE of prediction for enumerating live bacteria within the test set samples was 0.213.

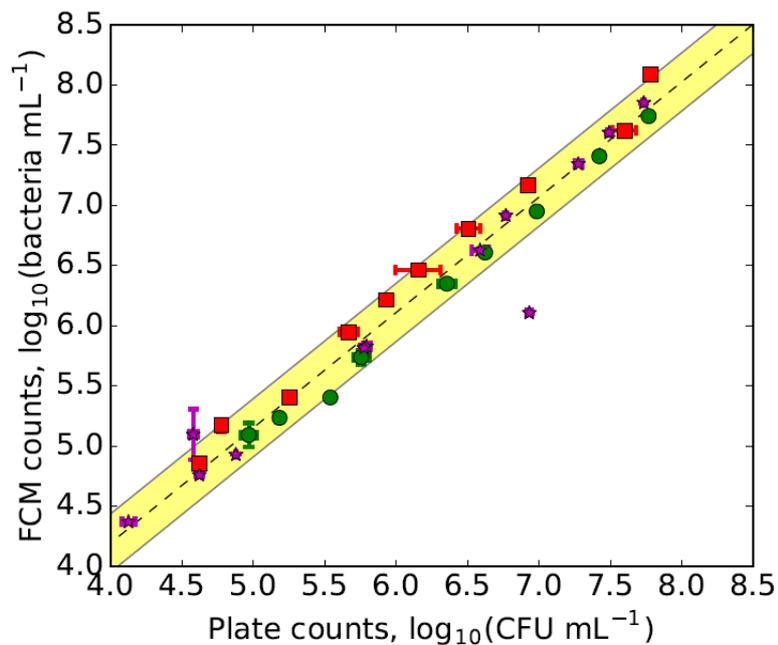

**Figure 4** Live bacterial concentration test set validation samples were analysed using the bead bottles 1 (●), 3 (■) and 5 (★). The dashed line is the standard curve of live bacteria and the shaded area represents the region

of plus or minus 2 standard errors (SE) of the standard curve line. The error bars in x and y directions of the data points indicate SE in replicate samples.

## Limit of detection for live bacteria

The limit of detection (LOD) of live bacteria in mixtures of live and dead was determined experimentally. As shown in Fig. 5, the FCM-based method is able to obtain absolute count of live *E. coli* having a concentration as low as $10^4$ bacteria ml$^{-1}$ before the measurements begin to deviate significantly from the standard curve. Although the current method cannot enumerate below $10^4$ live bacteria ml$^{-1}$, it can distinguish when the concentration is below this threshold.

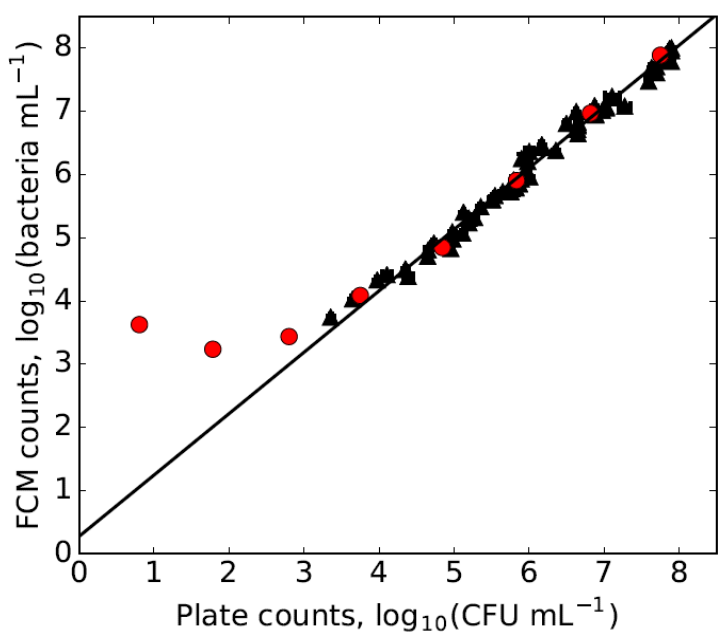

**Figure 5** Investigation of the limit of detection for the enumeration of live *E. coli*. To determine the lowest order of detectable concentration, live *E. coli* samples in 10 times dilution series (●) were analysed and plotted with the data used to obtain the standard curve (▲). The solid line represents the overall standard curve obtained for live *E. coli*.

## Standard curve for dead bacteria

The 'reference' concentration of dead bacteria is more difficult to obtain than that of live bacteria due to their non-culturability. To validate the enumeration of dead bacteria by FCM, it was compared to the estimation of dead bacterial cell count based on plate counting of input bacteria (obtained from steps A and B, Fig. 1). Fig. 6 shows that there is a linear correlation between the enumeration of dead *E. coli* obtained by the two methods (y = 0.996x + 0.062), which is not significantly different from a 1:1 line. The data were obtained by using three bottles of bead standards, on multiple days. Each data point (N = 41) on Fig. 6 is an average of triplicate measurements (n ≈ 123). $R^2$ for the overall standard curve of dead *E. coli* is 0.93 and the SE of linear regression is 0.296. SE associated with the expected concentration of dead bacteria includes errors associated with plate counts that were performed before the killing step (steps A and B in Fig. 1) as well as the errors induced by washing the sample. The error bars in the horizontal direction are significantly larger than the vertical direction, reflecting the superior precision of the FCM method.

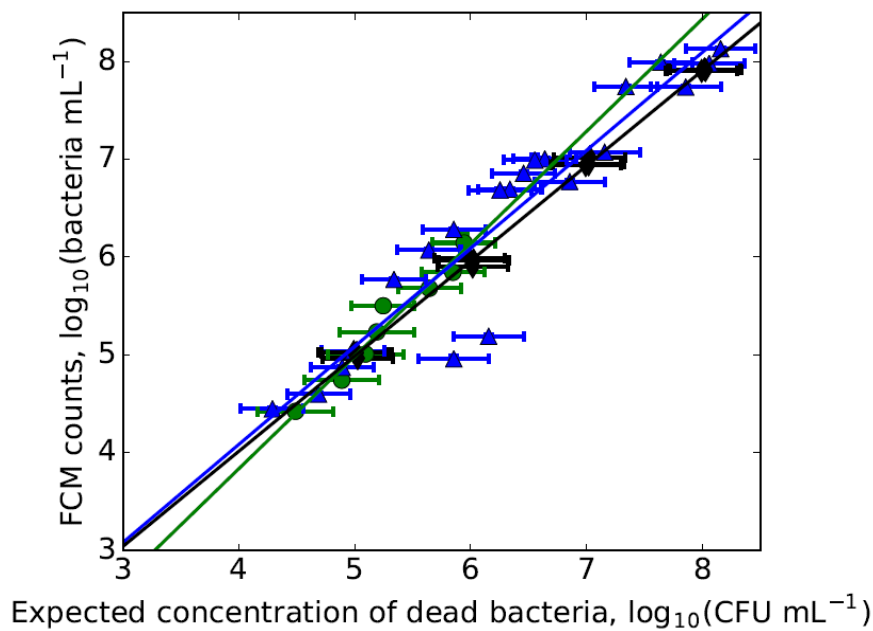

**Figure 6** Standard curve for the enumeration of dead *E. coli*. The concentration of dead *E. coli* enumerated by FCM plotted against the expected concentration of dead bacteria (as determined from plate counts at steps A and B in Fig. 1). Data were collected using different bottles of beads, numbered 1 (●), 2 (▲), and 4 (♦). The vertical error bars indicate the standard error in triplicate samples measured by FCM. The horizontal error bars

take into account the errors introduced by sample washing, and variations of plate counts from which the concentrations of dead bacteria were calculated.

## Validation for the standard curve of dead bacteria

To validate the FCM-based enumeration of dead bacteria, 22 test set samples were measured over 2 days and using different LIVE/DEAD Bacterial Viability and Counting Kits. Bead bottle 3 was used in the establishment of the standard curve of dead bacteria whereas bead bottle 5 was not. Bottle 5 was used to demonstrate that the standard curve does not need to be updated when using bottles of beads from batches different to those used in the establishment of the curve. Fig. 7 shows that with the exception of one outlier, regardless of the bottle of beads used the test set samples agree with the standard curve of dead *E. coli*. The data points begin to deviate slightly more from the standard curve at the lower concentrations, as the limit of detection ($10^4$ bacteria ml$^{-1}$) is approached.

The residual analysis on the test set data for enumerating dead *E. coli* showed that bottle 3 residuals were mostly above zero and bottle 5 residuals were below zero for the higher concentrations and above zero for the lower concentrations, consistent with the skew in Fig. 7. The SE of prediction for enumerating dead bacteria within the test set samples was 0.387.

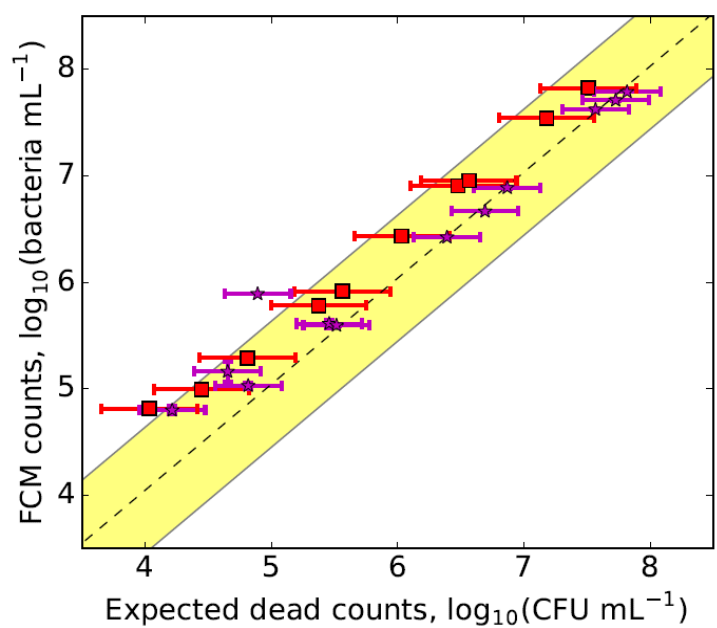

**Figure 7** Test set validation of dead bacterial concentration was completed using the bead bottles 3 (■) and 5 (★). The dashed line is the standard curve of dead bacteria and the shaded area represents the region of plus or minus 2 standard errors (SE) of the standard curve. The error bars in x and y directions of the data points indicate SE in replicate samples.

## Limit of detection for dead bacteria

To determine the range of concentrations of dead bacteria that the FCM method can enumerate, the LOD was found experimentally. As shown in Fig. 8, the absolute count of dead bacteria can be obtained for concentrations down to $10^4$ bacteria ml$^{-1}$ before the measurements begin to deviate significantly from the standard curve. Despite its inability to obtain absolute counts below $10^4$ bacteria ml$^{-1}$, the FCM method is able to distinguish when the concentration of dead bacteria is below this threshold. This yields a similar result to the limit of detection for live bacteria.

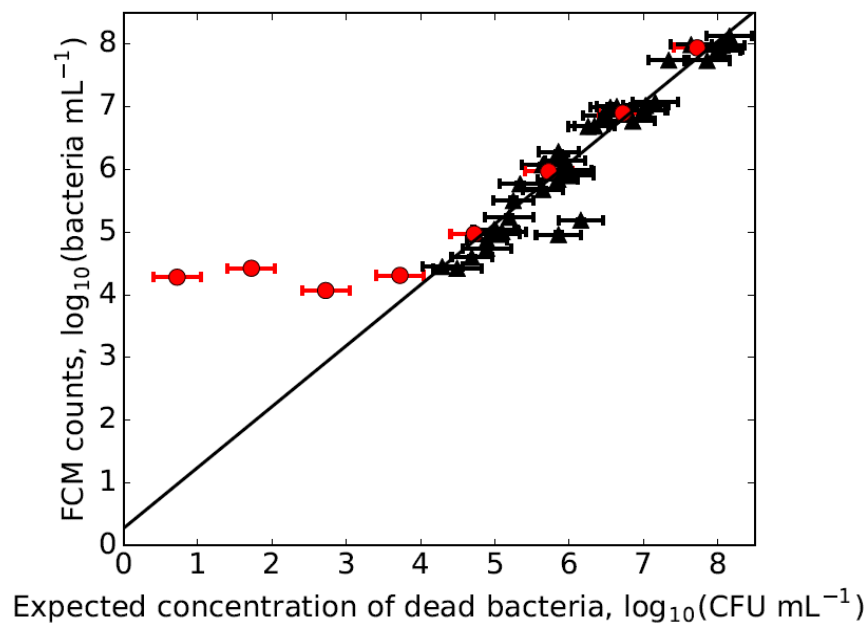

**Figure 8** Investigation of the limit of detection for enumerating dead *E. coli*. To determine the lowest order of detectable concentration, dead *E. coli* samples in 10 times dilution series (●) were analysed and plotted with the data used to obtain the standard curve (▲). The solid line represents the overall standard curve obtained for dead *E. coli*.

## Universal standard curve

The FCM counts of live and dead bacteria were based on two different fluorescence signals and calculated using the same formula (2). Therefore assuming the reference method (plate counting) was perfect, it is expected that the standard curves of live and dead bacteria are equivalent. To find an universal standard curve that can be used for the enumeration of both live and dead bacteria, both the live and dead counts obtained from FCM and plate count method were compared (Fig. 9).

The standard curves of live and dead bacteria were not significantly different. The linear regression obtained by combining the live and dead data was y = 0.970x + 0.260. Applying this universal standard curve to the live test set samples resulted in a SE of prediction of 0.216 which is worse than the result from using the standard curve of live bacteria. On the other hand, applying the universal standard curve to the dead test set samples resulted in a SE of prediction of 0.342 which is better than the result from using the standard curve of dead bacteria.

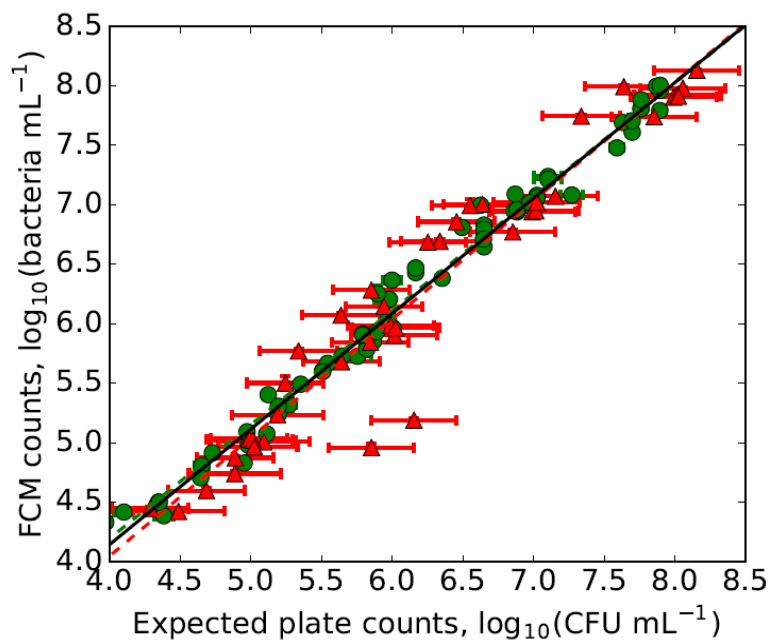

**Figure 9** The universal standard curve (solid line), based on the concentration measurements of both the live (●) and dead (▲) bacteria, in mixtures of live and dead. The standard curves of live and dead *E. coli* are illustrated by dashed lines. The error bars in x and y directions indicate the standard error in triplicate samples.

# Detection of live and dead bacteria in mixtures

To display the detection range and reliability of the FCM method, the FCM percentage, FCM counts of live and dead bacteria in mixtures, and their respective errors were plotted in Fig. 10. Each data point is the mean of triplicate measurements, which was compared to the expected concentration obtained using plate counts and the standard curve of the data set. In initial experiments the input plate counts for the dead bacterial suspensions were not collected. Hence no 'reference' bacterial concentration was available to determine the difference between some of the predicted and measured dead bacterial concentrations. More than 87% of the FCM counts of live bacteria and more than 84% of the FCM counts of dead bacteria were within 2 SE of the expected count.

Within a mixture of both live and dead bacteria, reliable enumeration of live bacteria was achieved when the live:dead concentration ratio varied from 100% down to *ca.* 2.5% live bacteria; and reliable enumeration of dead bacteria requires the percentage of dead bacteria to be between 100 and *ca.* 20%. A linear relationship exists between plate counts and FCM counts of live and dead bacteria ranging from $10^8$ down to *ca.* $10^4$ bacteria ml$^{-1}$.

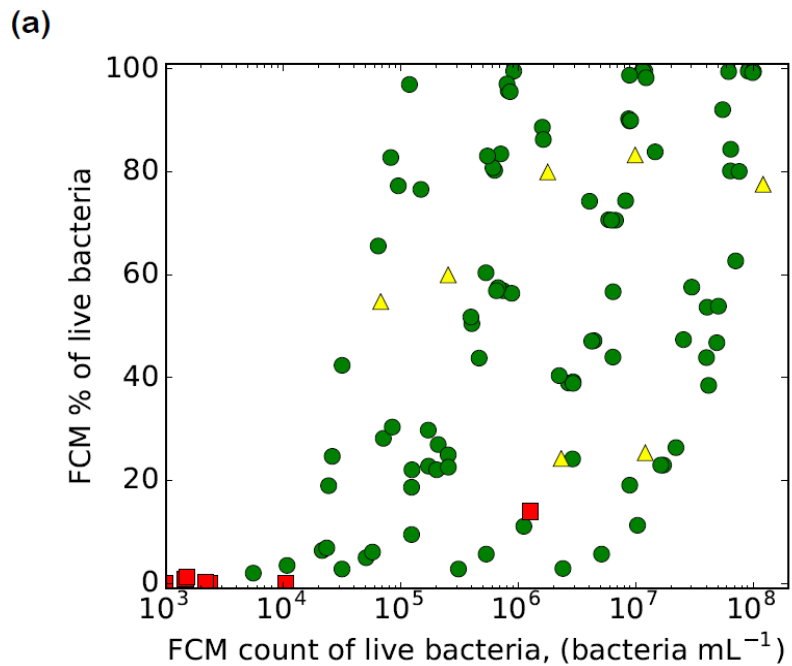

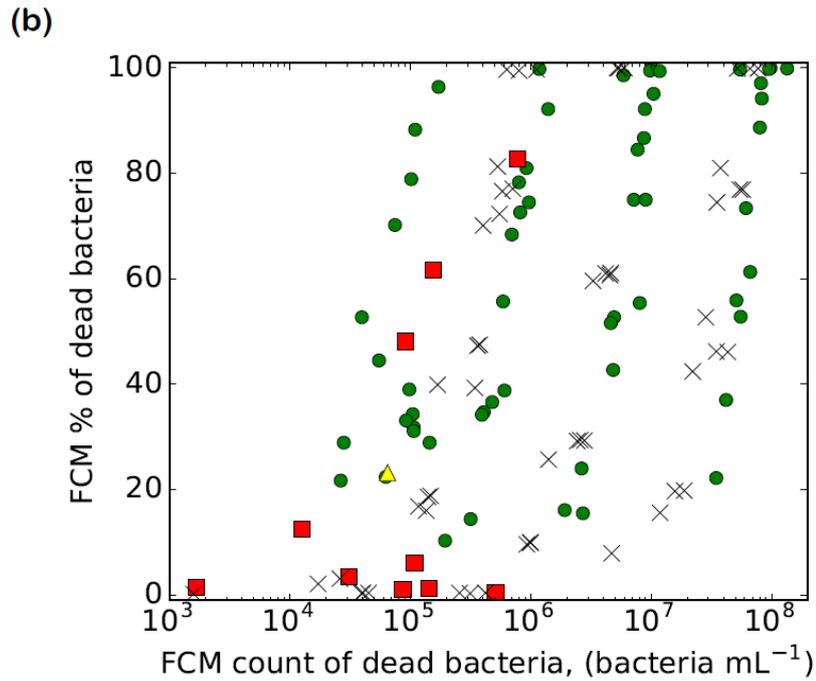

**Figure 10** The live (a) and dead (b) percentage and FCM counts collected from mixtures of live and dead bacteria. The standard error (SE) refers to the SE of linear regression for the respective standard curve data sets. In both the live and dead cases, the obtained FCM counts were compared to the expected FCM counts and each data point is labelled as either within 2 SE (●), over 2 SE but within 3 SE (△), more than 3 SE (■) or the error is not available (✕).

## Discussion

The FCM technique continues to become more commonplace in the field of microbiology. However, the ability to measure absolute concentration of bacterial cells is not a standard function on all FCM systems. In this study we have raised points of consideration when using the bead-based FCM counting method and applied the method to count varying ratios of live and dead *E. coli* for a wide concentration range of $10^4$ to $10^8$ bacteria ml$^{-1}$. The FCM-measured absolute counts correlated linearly with the concentrations obtained by the standard plate count.

## Bacterial concentration variation induced by sample preparation

Prior to comparing FCM counts with plate counts, it is important to check that the measurements are done on equivalent samples and that its concentration was not affected by sample processing in between. The plate counting done at different procedure steps showed that sample preparation induced variations in bacterial concentration. The bacterial population in samples are continuously varying due to cell growth and death. Population growth explains the often increased bacterial concentration following the incubation step. The variation in cell concentration from sample washing is more random and there are two factors that may contribute to this. The decrease in bacterial concentration may be caused by the loss of cells as they come loose from the pellet and poured out with the supernatant. On the other hand, it is possible that washing the bacteria may help reduce underestimation in plate counts caused by clumping.

The large variation in bacterial concentration shows the extent to which sample processing alters the bacterial samples before measurement by FCM. From this finding, one can no longer assume that the initial bacterial concentration is unchanged following sample processing; and that the concentration of dead bacteria cannot be determined based on the concentration of live bacteria, even if they underwent the same sample preparation steps. To obtain reliable enumeration of live bacteria, the sample was split into two, for measurement by FCM as well as for a final plate count to be done in parallel. The concentration of dead bacteria in the final samples were determined based on plate counting of the input bacteria prior to treatment with isopropanol, and taking into account the 28% mean absolute percentage deviation introduced by the sample washing steps.

## Vortexing of beads

The use of the bead standard was not straightforward. Prior to use, the bead solution must be homogenised to ensure the sample used in FCM experiments has the correct concentration of beads. In early trial experiments, the reference bead stock was sonicated for *ca.* 10 minutes then vortexed at 2200 rpm (*ca.* $24 \times g$) for 30 seconds before aliquoting into samples, subsequently the samples were vortexed prior to FCM measurements. This resulted in severe overestimation of the live cell concentration obtained by FCM, compared to the plate count results (Fig. S2). The problem was caused by the strong vortexing which induced foam formation

(Brando et al., 2000; Wulff et al., 2006). The introduced micro-air bubbles attracted the beads, and subsequently skewed the volume of beads pipetted (Brando et al., 2000; Wulff et al., 2006). Thereafter, mixing of beads were completed by gentle inversion of the sample and low speed vortexing at 5 rpm (*ca.* $0.00013 \times g$) to prevent foaming (Brando et al., 2000; Wulff et al., 2006).

The concentration of cells measured by the FCM method depend on the concentration of reference beads, hence it is important that the volume of beads added is accurate and consistent (Brando et al., 2000; Wulff et al., 2006). Variability in the pipetting of beads were tested for the standard curve experiments, by individually adding 10 µl of beads to each of the three analytical replicates of bacterial solution containing dyes. There were no major variations between these triplicate measurements and those obtained by measuring three analytical replicate samples. Thus, variation in the pipetting of bead solutions were negligible. In subsequent experiments, beads were added to each stained bacterial sample then divided into 3 aliquots for triplicate measurements.

## Enumeration of live bacteria

Viability is one of the key cellular properties investigated in microbiology, and is also one of the most debated topics (Roszak and Colwell, 1987; Kjelleberg, 1993; Servais et al., 1993; Barcina et al., 1997). Historically, viability has been defined by the cell's ability for reproductive growth and tested via the plate count method. However with different fluorescence techniques based on the detection of enzymatic activity, membrane integrity, transmembrane chemical and potential gradients, and gene expression, it is possible to differentiate beyond the classical definition based solely on cellular reproduction (Johnson, 2010; Zotta et al., 2012). In this study, SYTO 9 and PI were used, as cells stained by PI are unambiguously dead because PI cannot permeate through the cytoplasmic membrane. In our experiments, cells stained by SYTO 9 alone are identified as live because the cytoplasm is protected from the environment thereby preventing PI staining and allowing normal biological processes to take place. There is no universal dye combination for detecting viability, as the selection of fluorescent dyes depends on many factors including the cell physiology, conditions of the experiment and the measurement technique used (Zotta et al., 2012).

The proposed FCM method consistently counted approximately 0.273 $\log_{10}$(bacteria ml$^{-1}$) more live bacteria than plate counts. Other studies comparing FCM-based enumeration with plate counts also saw this phenomenon (Bensch et al., 2014; Buzatu et al., 2014). Moreover, the clumping problem was minimised in FCM due to its inherent ability for hydrodynamic focusing and by using a low flow rate of approximately 6 µl min$^{-1}$. In addition, a small subset of bacteria exists in a viable but non-culturable (VBNC) state, due to the stresses induced from sample preparation such as vortexing, staining and starvation. This small subset of VBNC bacteria is identifiable by FCM because its intact membrane retains SYTO 9 and prevents PI staining, but due to being non-culturable the plate count method cannot recognise them as live bacteria.

The FCM-based enumeration of both live and dead bacteria predicted well the respective plate counts. However we observed that for different bottles of beads there is a slight but consistent under- or over-estimation of the bacteria numbers. This consistency is observed regardless of whether bacteria are alive or dead, as demonstrated by the validation data obtained using bottle 3 for counting live (Fig. 4) and dead bacteria (Fig. 7). Data obtained using bottle 3, regardless of the type of bacteria it's used to count, results in a slight overestimation compared to the standard curve. The FCM method is dependent on the bead concentration and hence the enumeration obtained by using different bead bottles is skewed by the slightly different bead concentration of each bottle.

We demonstrated the analysis of live and dead bacterial mixtures using the FCM method, which was able to obtain absolute counts of the live and dead *E. coli* with a concentration ranging from $10^8$ bacteria ml$^{-1}$ down to approximately $10^4$ bacteria ml$^{-1}$. Below this limit, false-positive signals possibly due to interference from the electronic signal and sheath fluid of the flow cytometer became significant. Nonetheless, the method is able to distinguish when the concentration is below the $10^4$ threshold. Our results showed that the proposed method was able to reliably count live bacteria when the live:dead concentration ranged from 100% to *ca*. 2.5% live bacteria; and reliably count dead bacteria when the concentration ranged from 100% to *ca*. 20% dead bacteria.

The focus of the proposed FCM method was to enumerate total and subsets of bacteria with a total concentration ranging from $10^8$ to $10^5$ bacteria ml$^{-1}$. In the future, adjustments can be made to improve sensitivity that goes beyond the scope of the current study. For example, to detect low concentrations of bacteria the FCM procedures need to be adjusted by increasing the flow rate and duration of measurement, to enable the analysis of more cells i.e. a bigger volume at low concentration. Also, the amount of beads added to each sample would need to be decreased, so that the frequency of bead detection is not too high compared to that of the bacteria.

## Enumeration of dead bacteria

The concentration of dead bacteria cannot be determined directly from plate counts, neither can it be inferred from the final concentration of live bacteria even if they underwent the same preparation steps. This is because the concentration of bacteria was shown to be altered by the sample preparation process. Hence, to obtain a standard of reference for the dead bacteria concentration, plate counts were done before isopropanol was added (killing step, Fig. 1) and an expected concentration of dead bacteria was calculated (equation 1). We compared the FCM-measured concentration of dead *E. coli* to the plate count-based value of the expected dead concentration (Fig. 6). The errors associated with the expected dead bacteria concentration is considerable, because it accounts for the variations in plate count and the error introduced by the washing steps (step D, Fig. 1). Unlike the FCM method, plate counting requires dilution of the sample which introduce additional variations to the determination of the expected dead concentration. Nonetheless, the relationship between the dead bacteria concentration measured by the two methods is linear.

There are several advantages of enumerating dead bacteria using the proposed FCM method. Compared to the expected concentration calculated from plate counts at steps A and B (Fig. 1), the FCM method is faster, easier to carry out and has significantly smaller errors. In addition, the plate count method to obtain expected dead bacteria concentration cannot be used to analyse real life samples to determine how many dead bacteria are present. An alternative way to validate the enumeration of dead *E. coli* is to compare the proposed FCM method with using fluorescence microscopy. However as mentioned earlier, enumeration via microscopy is very laborious, time consuming, and is difficult to analyse large volumes of sample. On the other hand, FCM allows reliable and rapid measurements of large sample volumes which is advantageous in many applications, such as to monitor antibiotic efficacy.

## Universal standard curve

The standard curves of live and dead bacteria were not significantly different, which was expected as the FCM counts were calculated using the same formula. This also shows that there is no bias for the flow cytometer to count live or dead bacteria. The difference between the individual standard curves for live and dead bacteria is most likely due to the different methods of obtaining the reference plate count values. The method used to obtain the expected concentration of dead bacteria introduced large variations that were not present in obtaining the reference concentration for live bacteria. As most applications are focused on live bacterial detection (e.g. measuring bactericidal efficacy), it is important to preserve the predictive ability of the standard curve of live bacteria. Therefore the individual standard curves for live and dead bacteria are continued to be used instead of the universal curve.

## Detection of live and dead bacteria in mixtures

More than 84% of the FCM counts of live and dead bacteria were within 2 SE of the expected count. However this does not mean the precision of measuring live and dead bacteria are similar. Note that the SE of linear regression for the standard curve of live *E. coli* is 0.121, whereas that of dead *E. coli* is 0.296. The large error associated with the model of enumerating dead *E. coli* reflects the difficulty in obtaining reference counts of dead bacteria and that plate counting of input bacteria is not a precise way to achieve this.

Reliable enumeration of live and dead *E. coli* was obtained in the range of $10^8$ down to $10^4$ bacteria ml$^{-1}$. The investigated method is reliable for counting live *E. coli* when the proportion of *E. coli* concentrations range from 100% to 2.5% live; and reliable for counting dead *E. coli* when the concentration ranges from 100% to *ca*. 20% dead *E. coli*. The general bead-based FCM method has the potential to be applied in the measurement of bacterial concentration for mixtures of different fluorescently stained bacteria. It can be applied in situations where little information is known about the concentration or viability of the bacterial samples. The current study outlined detailed protocols and precautions for using the bead-based FCM method, and serves to lay the foundation for future analysis of different bacterial mixtures.

# Acknowledgements

We are grateful to the New Zealand Ministry of Business, Innovation and Employment for funding the *Food Safe; real time bacterial count* (UOAX1411) research programme. This work is a partial fulfilment of Fang Ou's PhD thesis, who is grateful for the University of Auckland Doctoral Scholarship and the Todd Foundation Award for Excellence. The authors thank Dr Julia Robertson, Janesha Perera and Stephen Edgar for their laboratory support.

# Conflict of interests

No conflict of interest declared.

# Supplementary material

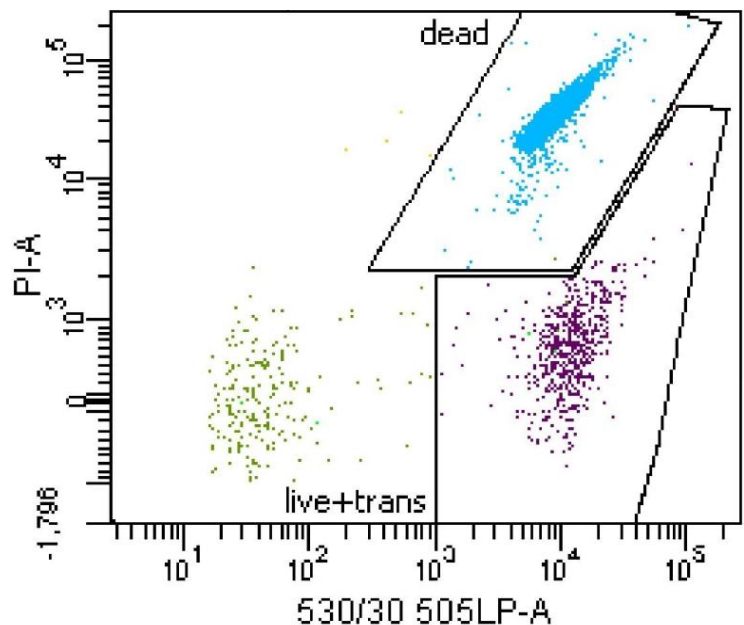

**Figure S1** A red fluorescence ('PI-A') versus green fluorescence ('530/30 505LP-A') flow cytometry (FCM) cytogram demonstrating the separation of live and dead bacteria for FCM counting analysis. The upper box labelled 'dead', frames the dead bacteria. Meanwhile the lower box labelled 'live + trans', frames the live bacteria and also includes those that are injured and stained by both SYTO 9 and PI. The unframed dots in the bottom left of the plot with low fluorescence intensity are considered noise and therefore excluded from analysis

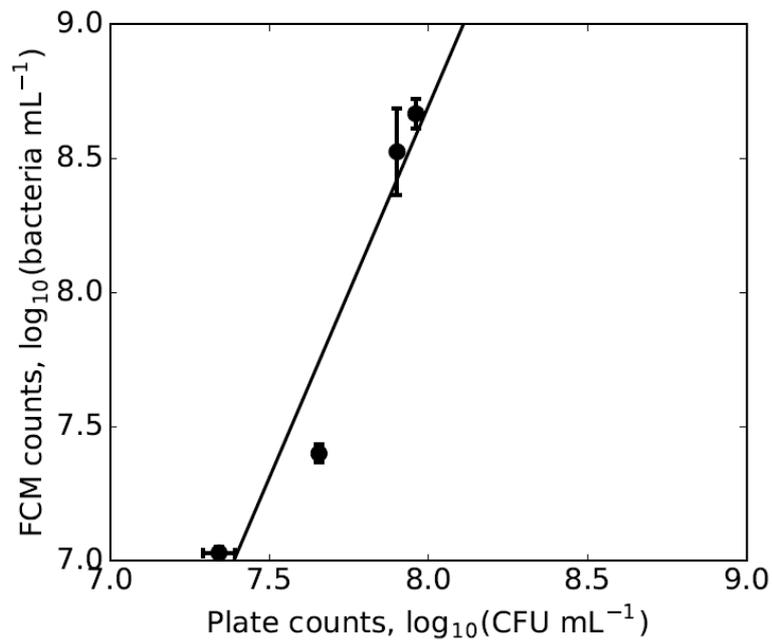

**Figure S2** Log of live *E. coli* concentration measured by FCM compared to that obtained by plate counts. FCM severely overestimated the concentration of live bacteria when the bead stock and samples were over-vortexed. The error bars in x and y directions indicate the standard error in triplicate samples.